\documentstyle[psfig,aps,prl]{revtex}
\begin{document}
\title{{\it A priori} Probabilities of
 Separable Quantum States}
\author{Paul B. Slater}
\address{ISBER, University of
California, Santa Barbara, CA 93106-2150\\
e-mail: slater@itp.ucsb.edu,
FAX: (805) 893-7995}

\date{\today}

\draft
\maketitle
\vskip -0.1cm

\begin{abstract}
\.Zyczkowski, Horodecki, Sanpera, and Lewenstein (ZHSL)
recently proposed a ``natural measure'' on the $N$-dimensional quantum
systems, but expressed surprise when it led them  to
conclude that for $N = 2 \times 2$, disentangled (separable)
systems are more probable ($0.632 \pm 0.002$) in nature
than entangled ones. We contend, however,
 that ZHSL's (rejected) intuition
has, in fact, a
sound theoretical basis, and that the {\it a priori}
 probability of disentangled $2 \times 2$ 
systems should more properly be viewed as (considerably)
less than 0.5.
We arrive at this conclusion in two quite distinct ways, the first
based on classical and the second, quantum considerations.
Both approaches, however, replace (in whole or part)
the ZHSL (product) measure by ones
based on the volume elements of {\it monotone} metrics, which in the classical
case amounts to adopting the Jeffreys' prior of Bayesian
theory.
Only the quantum-theoretic analysis --- which yields the smallest
probabilities of disentanglement --- uses the {\it minimum} number
of parameters possible, that is $N^{2} - 1$, as opposed to $N^{2} + N -1$
(although this ``over-parameterization'', as recently indicated by Byrd, 
should 
be avoidable).
However, despite
substantial computation,  we are not able to obtain precise
 estimates of these probabilities and
the need for additional (possibly supercomputer) analyses is
 indicated --- particularly
so for higher-dimensional quantum systems (such as the $2 \times 3$
ones, we also study here).

\end{abstract}


\vspace{.1cm}
\hspace{1.5cm} Keywords: Quantum entanglement, separable states, monotone
metrics, Jeffreys' prior, Haar measure,

\hspace{3.28cm} partial transposition

\vspace{-0.1cm}

\section{INTRODUCTION}

In a recent paper\cite{zycz1},
 \.Zyczkowski, Horodecki, Sanpera and Lewenstein (ZHSL)
\cite{zycz1} sought to 
estimate ``how many entangled (disentangled) states exist among all
quantum states''.  They gave three principal reasons for their
study: (1) to answer the question ``is the world {\it more classical}
or {\it more quantum}?''; (2) to know, for the purposes of numerical
simulation, ``to what extent entangled quantum systems may be considered
as typical''; and (3) ``to investigate how frequently certain  
nonseparable states, 
`peculiarly' admitting time reversal in one subsystem,
arise''. 
In response to the first query, ZHSL
concluded --- to their ``surprise'' \cite[p. 889]{zycz1} --- that
 although the 
(higher-dimensional) ``world'' is, in general, more quantum than
classical, this is not so for the $2 \times 2$ quantum systems.
We contend here, however, that alternative 
 analyses based on the concept
of {\it monotone} metrics on classical and quantum systems
\cite{petzsudar}, lead to the elimination of
this exception to their general rule.

In their investigation, ZHSL obtained a variety of both
analytical and numerical bounds on the volumes of the sets of separable
states for various dimensions,
 using what they asserted was a ``natural measure'' on the space of
density matrices (cf. \cite{wootters}).
 In the first analytical part of this
 communication (sec.~\ref{semiclassical}), we
indicate
an essential degree of arbitrariness in the choice of measure by
ZHSL, and its consequences for
the results they have reported (cf. \cite{slater1}).
We then argue  in favor of a specific 
alternative --- well-founded on  statistical
principles --- which leads to a markedly {\it smaller} probability of
encountering a disentangled (separable) state. (The numerical
results we report are for the $2 \times 2$,
$2 \times 3$ and $3 \times 3$ quantum systems.)
Then, in sec.~\ref{quantum},
 we study the use of methods more 
fundamentally quantum-theoretic  in nature --- requiring
us to develop 
a quite distinct set of procedures than those used by ZHSL and
followed in sec.~\ref{semiclassical}. (Due to the associated
  large 
 computational demands, we have  primarily limited our analyses to
the $2 \times 2$ systems, but in sec~\ref{quanttwobythree} we do, in fact,
initiate a parallel investigation of the $2 \times 3$ systems.)
We obtain for each of more than thirty   pairs of parameters, determining
the fineness of  approximating square grids
and three-dimensional simplicial decompositions, a set of three
 probabilities of disentanglement (Table~\ref{table1}), each
probability being 
based on a distinct form of
monotone metric \cite{slater2,petztoth}. The  sets are intended to
determine   a range of values within which any suitable candidate 
(meeting underlying natural criteria of {\it monotonicity}) for the ``true'' 
{\it a priori} probability of disentanglement
must lie. Essentially all such  probabilities we obtain turn
out to be considerably smaller 
 than both the result of ZHSL ($0.632
\pm .002$)
 and the alternative  to it ($\approx 0.35$)
 we promote in sec.~\ref{semiclassical}, based on the Jeffreys' prior
of Bayesian theory.
However, the need for additional computational work is
 indicated in order
to sharpen   the estimates reported in 
secs.~\ref{n123n27} \ref{n119n28} and \ref{additional},
 as well as to extend our general approach to
higher-dimensional quantum systems.
(We are reminded, to some degree, of the computational/combinatorial
 challenges
of lattice gauge theory \cite{davies}.)
In sec.~\ref{restricted}, we switch from the  explicit enumeration 
approach (based on regular grids and simplicial decompositions) to a
randomization methodology  (such as ZHSL employed in their studies) --- but
 also find this to be highly computationally demanding, since we must
search 
in a high-dimensional parameter
 space for those particular points corresponding to density matrices.

A conservative evaluation of the accumulated evidence of the multiple 
quantum-theoretic analyses we report (Tables~\ref{table1} \ref{table2}
\ref{tablerandom}) indicates that
the {\it a priori} probability of disentanglement 
for the $2 \times 2$ systems should be regarded --- using
any of a continuum of possible
 acceptable standards, in particular that provided by the
{\it minimal}  monotone (Bures) metric --- as
{\it no more} than 
eleven percent. As to a {\it lower} bound, on the other hand,
 on the probability of separability,
it remains an unsettled issue as to whether or not the
{\it maximal} monotone metric should be
viewed as furnishing a bound  strictly greater
than zero.

In our concluding remarks (sec.~\ref{Concluding}), we draw attention to
an interesting 
 recent analysis of M. Byrd (personal communication), bearing upon the
issue of whether or not the use of ``over-parameterizations'' by ZHSL
and (following them) by us in sec.~\ref{semiclassical}, can be averted.

\section{SEMICLASSICAL STATISTICAL ANALYSES OF $3 \times 3$, $2 \times 3$
and $2 \times 2$ QUANTUM SYSTEMS} \label{semiclassical}
ZHSL \cite{zycz1}
 used as a measure on the space of $N \times N$ density
matrices the product of the Haar measure for the unitary group
$U(N)$ and the uniform distribution on the $(N-1)$-dimensional
simplex spanned by the $N$ eigenvalues of the density matrix.
Now, we see no basis (within the
semiclassical framework adopted by ZHSL) for questioning the
use of the Haar measure.
However, the selection of  the uniform distribution 
on the simplex appears not to be
so compelling, as it lacks as convincing  a rationale as the group-theoretic
argument
for the Haar measure (cf. \cite{boya}).
Also, we must point out that the analyses of ZHSL are based on
``over-parametrizations'', since $N^{2} +N -1$ parameters are used,
while the convex set of $N \times N$ density matrices is only
$(N^{2} -1$)-dimensional in nature.
Though we adhere to this over-parameterization in the analyses
of this section, in sec.~\ref{quantum}  we revert to the more
natural and conventional form.

The uniform distribution  on the $(N-1)$-dimensional
simplex ($p_{1} + \ldots + p_{N} = 1$; $p_{i} \geq 0$)
can be considered to be that specific
 member of the (continuous) family of Dirichlet probability 
distributions \cite[sec. 7.7]{wilks,ferguson},
\begin{equation} \label{dirichletdist}
{\Gamma(\nu_{1} + \ldots + \nu_{N}) \over \Gamma(\nu_{1}) \ldots
\Gamma(\nu_{N})} p_{1}^{\nu_{1} -1} \ldots p_{N-1}^{\nu_{N-1} - 1}
(1- p_{1} - \ldots - p_{N-1})^{\nu_{N}-1}; \quad
\nu_{1} > 0, \ldots, \nu_{N} > 0
\end{equation}
which has all its $N$ parameters ($\nu$'s) set equal to unity.
The family of Dirchlet distributions
 is {\it conjugate}, in that if one selects a 
{\it prior} distribution belonging to
 it, then, through the application of Bayes' rule
to observations drawn from a multinomial distribution, one arrives
at a {\it posterior} distribution which is also within the family.

\subsection{Jeffreys' prior}
Of strongest interest, however, for our purposes  here, is that the principle
of reparameterization invariance (based on the {\it Fisher information}
\cite{frieden1,frieden2}) leads to
the special case
(the Jeffreys' prior) in which (\ref{dirichletdist}) has  all its
 $N$ parameters
set equal to {\it one-half} \cite[eq. (3.7)]{berger}
 \cite{kass} and not {\it unity},
 as for the uniform distribution.
``The main intuitive motivation for Jeffreys' priors is {\it not} their
invariance, which is certainly a necessary, but in general far from
sufficient condition to determine a sensible reference prior; what makes
Jeffreys' priors unique is that they are {\it uniform} measures in a
particular metric which may be defended as the `natural' choice for
statistical inference'' \cite{bernardokass}.
By way of
illustration,
 Kass \cite[sec. 2]{kass} (cf. \cite{antonelli}), using the transformations
$p_{i} = 2 z_{i}^2$,  demonstrates how the Jeffreys' prior  for the
trinomial model ($N=3$) on the two-dimensional simplex, can
be (making use of spherical polar
coordinates) transformed to  the uniform distribution
on the positive-octant portion of the two-dimensional
 sphere, $z_{1}^{2} + z_{2}^{2} + z_{3}^2 = 4$  of radius 2.
(Braunstein and Milburn \cite{braunstein} show that for two-level {\it quantum}
systems, statistical distinguishability is just the [Bures/minimal
monotone] metric
on the surface of the unit sphere in {\it four} dimensions.
In contrast, the space of $n$-level quantum systems is ``not a space
of constant curvature for $n > 2$
 and not even a locally symmetric space. The physical
meaning of this fact seems to be an interesting open question''
\cite{dittmann} (cf. \cite{dittmann2,hiai}).)

Clarke and Barron \cite{clarke1,clarke2} (cf.
\cite{rissanen}) have established
 that Jeffreys' priors 
(the normalized
volume elements of Fisher information metrics)
asymptotically maximize Shannon's mutual information between
a sample of size $n$ and the parameter, and that Jeffreys' prior
is the unique continuous prior that achieves the asymptotic minimax
risk when the loss function is the Kullback-Leibler distance between
the true density and the predictive density. (The possibility
of extending the ``universal coding'' results  of Clarke and Barron
 to the {\it quantum} domain, has been investigated
in \cite{kratt} (cf. \cite{jozsa}).)
Clarke \cite{clarke3} asserts that ``Jeffreys' prior can be justified by
four distinct arguments''. In addition, 
Balasubramanian \cite{vj} ``cast parametric model selection as a disordered
statistical mechanics on the space of probability distributions'' and 
``derived and discussed a novel interpretation of Jeffreys' prior as the
uniform prior on the probability distributions indexed by a
parametric family'' (cf. \cite{vj2}).

Now, it is of interest to note that in the limit in which  the $N$ parameters
($\nu$'s)
of the Dirichlet distribution (\ref{dirichletdist}) all
go to zero, the distribution becomes totally
concentrated on the pure states of the $N$-dimensional quantum system.
Since ZHSL showed that
in  ``the subspace of all pure states, the measure of separable
states is equal to zero'' \cite[p. 886]{zycz1}, we would anticipate, making
use of a continuity argument, that the measure or 
 volume of the set of separable states would increase if all
$N$ parameters of (\ref{dirichletdist}) were  fixed at one-half
(Jeffreys' prior), but still be less than if they were all taken to be
equal to unity, {\it etc}.
(``The purer a quantum state is, the smaller its probability of being
separable'' \cite[p. 891]{zycz1}.)
\subsection{The case of $3 \times 3$ quantum systems} \label{sec3x3}
We have, in fact, tested these last
 contentions regarding competing measures of
separability, through numerical means, first, 
generating a set of three
 thousand random $9 \times 9$ unitary matrices ($N=9$),
following the (Hurwitz/Euler angle)
 prescription given in \cite[eqs. (3.1)-(3.5)]{zycz2}.
From it, we produced (in the manner
of ZHSL 
\cite[eq. (34)]{zycz1}) three sets of three thousand  $9 \times 9$
density matrices: one set based on the selection
$\nu_{1} = \ldots = \nu_{9} ={1 \over 2} $; another for $\nu_{1} = \ldots =
\nu_{9} = 1$ (as, in effect,  done in \cite{zycz1}); and a third for
$\nu_{1} = \ldots = \nu_{9} = {3 \over 2}$.
(Random realizations of the Dirichlet distributions were generated
based on the fact that they can be considered to be joint distributions of 
[univariate] gamma distributions \cite{wilks,ferguson}.
The three thousand instances we obtain
 are  obviously far fewer in number than 
the ``several millions''
ZHSL \cite{zycz1} apparently employed as a general rule
in their series of analyses of quantum systems of various dimensions.
 This is primarily due to our
full reliance on MATHEMATICA, while ZHSL  --- as K. 
\.Zyczkowski wrote in a personal communication --- employed FORTRAN routines
for  random number generation. Nevertheless, as noted immediately below,
 \.Zyczkowski, using his speedier
 routines, has confirmed the main aspects of our analysis.
Additionally, the {\it quantum-theoretic} analyses of sec.~\ref{quantum}
require a quite different set of algorithms, and it is far from clear
whether our use there of MATHEMATICA is in any way relatively inefficient.)

Then, we determined  whether all (nine of) the eigenvalues  
of the {\it partial transpositions}
of the random density matrices (viewing
them  as $(3 \times 3) \times (3 \times 3)$
density matrices, in the manner
of (21) in \cite{horo}) were positive (as they must be in the separable
case) or not.
For the $\nu = {1 \over 2}$ 
(Jeffreys' prior) scenario,  eighty-three of the three thousand
density matrices had this 
positivity property, while considerably more (six hundred and two)
 possessed it 
for $\nu = 1$ and   still more (one thousand two hundred and ninety-six)
 for $\nu = {3 \over 2}$. Thus, we note an approximate decrease by a factor of
${83 \over 602} \approx .138$ in the 
upper bound on our suggested probability of encountering a separable
state {\it vis-\'a-vis} the analysis of ZHSL.
\subsection{The cases of $2 \times 3$ and $2 \times 2$ quantum sytems} \label{semi2by3}
We, then, conducted parallel analyses to those in sec.~\ref{sec3x3}
 for the $2 \times 3$
and $2 \times 2$ systems. (For both such systems, but not
higher-dimensional ones, such as the $3 \times 3$, the positivity of
the partial transposition is a sufficient, as well as necessary condition
for separability \cite{horo}. So, we will be estimating probabilities
themselves, rather than upper bounds on them.) In both cases,
 we now employed ten thousand
realizations. In the $2 \times 3$ case, we found 1,309 separable states, using
$\nu = {1 \over 2}$,  and 4,135 for $\nu= 1$,
 as well as  6,357 for $\nu = {3 \over 2}$.
(Our statistic of .4135 needs to be compared with that of $.384 \pm
.002$ of ZHSL --- which, as noted, was based on a much larger sample.)
For the $2 \times 2$ systems, the analogous results were: 3,633; 6,564;
and 7,946.
So, we would conclude, in this analytical framework,
 that the proportion of separable states among the
$2 \times 2$ quantum systems should be taken to be
 approximately .36 --- which is well {\it below}
 the 
demarcation point of .5, {\it above}
 which ZHSL found their result of $.632 \pm .002$ 
(roughly comparable  to ours of .6564) to
(counterintuitively) lie.

K. \.Zyczkowski has kindly
repeated the analyses reported above for the case $\nu = {1 \over 2}$
(that is, Jeffreys' prior), using 200,000 random realizations for each
of the three scenarios. The probabilities of separability he obtained were
(all digits being significant he states): 
.022 (for the $3 \times 3$ systems); .122 ($2 \times 3$ systems);
and .352 ($2 \times 2$ systems). These should be compared with our
results
(based on considerably smaller 
samples) of .0277, .1309 and .3633, respectively.

These various numerical results are, thus, quite supportive of our
arguments and help to fulfill the first objective of  this letter of showing
the dependence of estimates
 of the volume of the set of separable states
on the particular choice of (symmetric) Dirichlet distribution on
the $(N-1)$-dimensional simplex spanned by the $N$ eigenvalues of
the $N \times N$ density matrix ($\rho$).
We note again that ZHSL \cite[p. 889]{zycz1}
expressed ``surprise that the probability that a mixed state 
$\rho \in H_{2} \times H_{2}$ is separable exceeds fifty percent''.
Thus, they  would have apparently been not so confounded if the uniform
distribution on the three-simplex of eigenvalues had been
replaced by the Jeffreys' prior, since its use yields a {\it more
modest} percentage of
approximately thirty-five.

Let us also note that our suggested modification ($\nu = {1 \over 2}$)
 of the ZHSL measure  ($\nu =1$) would appear to find some support
in a recent paper concerned with a parameterization (sharing certain
features with that of
ZHSL)
of the $N \times N$ density matrices \cite{boya}. 
Its authors consider the $N$ eigenvalues to be parameterized by the 
``squared components'' of the $(N-1)$-sphere
(rather than coordinates in the $(N-1)$-dimensional simplex, as in ZHSL).
As noted above, in relation to \cite{kass}, the Jeffreys' prior is simply
the uniform distribution on such a {\it sphere} --- while ZHSL used 
instead the uniform
distribution on the {\it simplex}.

\section{QUANTUM-THEORETIC 
 STATISTICAL ANALYSES OF $2 \times 2$ AND $2 \times 3$ SYSTEMS} \label{quantum}
In \cite[sec. II.C]{slater1},
 we presented evidence that certain statistical features
of the product measure employed by ZHSL \cite{zycz1} were not
reproducible through the use of any of the possible (continuum of)
{\it monotone} metrics.
 A similar conclusion appears to hold if one replaces the
uniform distribution in the ZHSL product measure 
(as we have done above in sec.~\ref{semiclassical}) by any other member of the
family of Dirichlet distributions
(\ref{dirichletdist}). Since Petz and Sud\'ar \cite{petzsudar}
have argued that monotone metrics are the quantum analogues of
the (classically unique) Fisher information metric,
it would seem highly desirable to replace the product measures
so far employed  
by ones based directly on the volume elements of such metrics.
(In \cite{slatertherm}, efforts were reported to integrate the volume
elements of the {\it minimal} and {\it maximal}
 monotone metrics over the convex sets
of $3 \times 3$ and $4 \times 4$ density matrices.)
In so doing, we would avoid the  nonparsimonious
``over-parameterization'' mentioned at the
outset of sec.~\ref{semiclassical}.
(However, it will be incumbent upon us to develop a quite distinct
set of computational methods than those used by ZHSL and applied in
sec.~\ref{semiclassical}.)

We have, in fact, conducted such a series of analyses for the $2 \times 2$
quantum systems, based on a MATHEMATICA program containing {\it two} parameters
of choice,
$n_{1}$ and $n_{2}$. The parameter $n_{1}$ determines the fineness of a
regular decomposition of the three-dimensional simplex --- the 
points of which correspond now
to the {\it diagonal} entries of $\rho$, and not the {\it eigenvalues}, as in 
sec.~\ref{semiclassical} and 
the work of ZHSL
\cite{zycz1} and Boya {\it et al}
\cite{boya}. (Of course, both the eigenvalues and diagonal entries of a
density matrix are nonnegative and sum to unity.
 To compute the coordinates of the simplicial 
coordinates, we followed
an algorithm for the next {\it composition} of an integer $N$ into $K$ parts,
given
in \cite[chap. 5]{nij}, taking $K=4$ for our
purposes, and then dividing each of the
${N + 3 \choose N}$ compositions generated by $N$.) The reciprocal of the
 parameter $n_{2}$ is the distance between adjacent points
of a regular square
grid --- having its extreme points/corners  at $({1 \over 2},{1 \over 2}),
({1 \over 2},-{1 \over 2}),(-{1 \over 2},-{1 \over 2})$
and $(-{1 \over 2},{1 \over 2})$ --- imposed on a circle of radius one-half
centered at the origin of the complex plane. (The off-diagonal
entries of a density matrix can not exceed one-half in absolute value,
so they must lie within this circle.)

For
 specific values of  $n_{1}$ and $n_{2}$, within but challenging our
computational capabilities, we generated  the
associated 
three-dimensional simplicial decompositions
 and twelve-dimensional uniform lattices
(the six-fold Cartesian product of the imposed two-dimensional
 square grid --- six, of course, corresponding to the number of pairs of
off-diagonal entries). Then, we explicitly enumerated all those points in the 
fifteen-dimensional product space
 parameterizing the $4 \times 4$ density matrices of mixed states
(that is, yielding matrices having all strictly positive eigenvalues,
noting that the additional Hermiticity and trace requirements are
automatically  satisfied by
construction). We would reject any density matrices of pure states
(the totality of which form a six-dimensional subspace \cite{vanik})
that happened to be generated, since our measures
(see immediately below) are singular on them, as well as more generally,
degenerate density matrices, those density matrices not being of full rank
(and hence having zero determinant).
However, the possibility remains --- in particular, since we will be
computing (nonrobust) {\it averages} --- that the 
behavior of the measures for a relatively few {\it nearly} degenerate
states,  can strongly influence the  results
(cf. Tables~\ref{table1} and \ref{table2}).

 By our purposeful design, the explicitly enumerated points are uniformly
distributed (using the conventional parameterization)
 in the fifteen-dimensional convex set of $4 \times 4$ 
density matrices.
We took several significant
 steps in our MATHEMATICA  (``backtrack''  \cite[chap. 27]{nij})
 program to cut down on the
(potentially huge)
search spaces,  by utilizing the requirement that all the principal minors of
a density matrix must be nonnegative \cite[Thm. 7.2.5]{horn}
 \cite{bloore}. 
(A {\it sufficient} condition only, of possible interest, would be that
 the matrix is {\it diagonally-dominant} \cite{horn}.)
Also, in our later, larger analyses, we exploited certain permutational
symmetries.

We, then, employed an {\it ansatz}  of ours \cite{slatertherm},
 building upon a result of Dittmann
\cite[p. 76]{dittmann}
(pertaining to the Bures or minimal monotone metric) regarding
the spectrum of the sum of the operators of left and
right multiplication of matrices cf. \cite[p. 112]{bhatia}. Utilizing 
 it, we assigned as a weight to each
density matrix  ($\rho$) generated (the eigenvalues of which are denoted by
$\lambda_{i}$), 
the volume elements of certain  monotone metrics
of particular interest. These elements we took
to be of the form $[\Pi_{i,j=1}^{4} f(i,j)]^{{1 \over 2}}$, where 
(the ``Morozova-Chentsov'' function \cite{petzsudar}
\cite[eq. (3)]{petzlaa}) $f(i,j)$ is equal to 
${2 \over (\lambda_{i} +\lambda_{j})}$ in the minimal monotone case,
$(\lambda_{i} + \lambda_{j}) \over 2 \lambda_{i} \lambda_{j}$ in
the maximal monotone case, and for the Kubo--Mori/Bogoliubov (KMB) metric
(associated with the relative entropy)
 \cite{petztoth,petzkubo1,balian},
${\log{\lambda_{i}} - \log{\lambda_{j}} \over 
\lambda_{i} -\lambda_{j}}$. (In this last case, if $\lambda_{i} = \lambda_{j}$,
 we take
$f(i,j) = \lambda_{i}^{-1}$.
It is interesting to note that the inverses of these
 Morozova-Chentsov functions
are simply well-known indicators  of {\it central tendency},
 such as the arithmetic
mean, the logarithmic mean, and the harmonic mean \cite{qi}. Choosing a particular
monotone metric is, therefore, akin to selecting such an indicator.)
We also
checked if $\rho$ satisfied the partial transposition condition,
necessary for separability.

We now report several  sets of results in this regard --- but
let us first make some important observations
(taking into account
 that the determinant of a matrix is equal to the product of its
eigenvalues).
Using the formulas just given, one can show \cite{slatsqueezed} that 
for an $N \times N$ density matrix ($\rho$), the volume element
of the minimal monotone metric (and also the KMB-metric)
 is directly proportional to 
$(\det{\rho})^{-{1 \over 2}}$, while for the maximal monotone metric,
the volume element is directly proportional to $(\det{\rho})^{{1 -2 N  \over 2}}$.
So, the divergence near the  boundary of degenerate states
($\det{\rho}= 0$) of the volume element of
the maximal monotone metric is much more severe than for the other two
metrics 
under investigation.
In fact, in our previous studies
\cite{slater2,slatertherm}, we have concluded that the integral
of the volume element of the maximal monotone metric over the convex set
of $2 \times 2$ density matrices does {\it not} converge
(in contrast to those for the minimal monotone and KMB-metrics).
(For $N > 2$, however, the 
issue of convergence appears to be unsettled.)
 So, it would seem  that --- unless one chooses to remove from consideration
(as was done in \cite{slater2}, for inferential purposes) those 
states the degeneracy of which exceeds some prescribed level
\cite{slater2} (cf. sec.~\ref{truncate}) --- one
 can not, in fact, define a probability
distribution based on the maximal monotone metric. We have been
able, however, for $N=3$, by taking the limit of a certain ratio, to obtain
associated {\it marginal} probability distributions using
 the maximal monotone metric \cite{slatertherm}.) 
Also, the maximal monotone metric is of substantial
 interest, in that it has been
characterized as the most {\it noninformative} of the monotone metrics
 \cite{slater2,petztoth}.

The motivating  hypothesis for pursuing the analyses immediately below is that
one should be able to find values of 
the parameters $n_{1}$ and $n_{2}$ large enough
(say, $m_{1}$ and $m_{2}$), so that the associated  probabilities of 
disentanglement are within some $\epsilon$ of each other for
{\it any} choices of $n_{1}$ and $n_{2}$
which {\it dominate} both $m_{1}$ and $m_{2}$.
(It is useful to bear in mind, however, that there is a qualitative
difference between analyses based on {\it even} or 
{\it odd}  values of $n_{2}$, as
will be indicated.)
This would indicate a convergence
of these probabilities
in the continuum limit  as $n_{1}$ and $n_{2}$  
each become indefinitely large.

\subsection{The case $n_{1} = 23$, $n_{2}=7$} \label{n123n27}

The choice of $n_{2}=7$ leads to a square grid, having
 thirty-two points --- serving as trial off-diagonal entries --- lying
within the circle of radius one-half.
(The particular arrangement of the lattice points, then,
 mandates that those density matrices we will be able to construct
will have off-diagonal entries of modulus no less than
${1 \over 7 \sqrt{2}} \approx .101015$. This, in turn, implies that the
product of any pair of diagonal entries of the
density matrices will not be less than this value.)
The numbers of density matrices we were, then, able to construct
were 1,340,928. Of these, 356,096 
passed the transposition test for separability.
Applying the weights based on the three  monotone metrics
considered, we obtained prior
probabilities of encountering separable states of
\begin{equation} \label{probs1}
p_{min} =.111102 ,\quad p_{KMB} = .0873186,\quad p_{max} = .0846153.
\end{equation}
Of course, these three values are
 all considerably less than both the ZHSL statistic of 
$.632 \pm .002$ and the preferred one (of the three given)
 of sec.~\ref{semiclassical}
based on the Jeffreys' prior, that is, $\approx .35$.

We have also computed the ``degree of entanglement'', $\sum_{i=1}^{4}
\lambda_{i}^{'} - 1$ (which must lie between zero and unity) for all the (1,340,928) density matrices and 
averaged the results  with the same set of three weights as used to obtain
(\ref{probs1}). The outcomes were
\begin{equation}
d_{min} = .18206, \quad d_{KMB} = .208022, \quad d_{max} = .248457.
\end{equation}
The corresponding value obtained by ZHSL for the $2 \times 2$ systems was
considerably smaller, that is .057
\cite[App. B]{zycz1}.
(ZHSL remarked that this quantity seemed to saturate at approximately 0.10
for large systems.)

\subsection{The case $n_{1}= 19$, $n_{2}=8$} \label{n119n28}

Since the parameter $n_{2}$ is now an even
integer,  the origin 
(0,0) of the complex
plane becomes, by our particular mode of 
construction, one of the forty-nine intersection
 points of the square grid lying within
the circle of trial values for the off-diagonal entries.
There are, then,  no nontrivial lower bounds imposed on the moduli of
these entries, as there are for odd values of $n_{2}$ --- such
as {\it seven} in the immediately preceding analysis of sec.~\ref{n123n27}.
We obtained 4,443,408 density matrices, of which
1,284,816 satisfied the separability criterion. Use of the volume elements
of the three selected monotone metrics as weights resulted in
\begin{equation} \label{probs2}
p_{min}= .147968, \quad p_{KMB} = .123283, \quad p_{max} = .0554999,
\end{equation}
and
\begin{equation}
d_{min} = .18686,\quad d_{KMB} = .184149, \quad d_{max} = .17713.
\end{equation}

\subsection{Additional (nontruncated) analyses} \label{additional}

Continuing along the same lines as secs.~\ref{n123n27} and \ref{n119n28},
we have conducted analyses for additional choices 
 of $n_{1}$ and $n_{2}$.
(It is interesting to note that unit increases in $n_{2}$ are relatively more costly
computationally than in $n_{1}$.)
We report our accumulated set of results 
in Table~\ref{table1}. The analyses are listed in
 increasing order of the total number of density
matrices generated. (During the course of conducting these analyses, we were
able to undertake larger-sized studies, corresponding to those listed at
the bottom of the table, by taking advantage of certain
inherent permutational symmetries. An analogous assertion can be made in 
regard to Table~\ref{early}.)

As a general rule, the probability of disentanglement is greatest for
the minimal monotone metric, although still markedly less than the ZHSL
result ($0.632 \pm 0.002$) or that of sec.~\ref{semiclassical}
($\approx 0.35$) based on the Jeffreys' prior.
The stability of the results, on the other hand, is least for the maximal
monotone metric (in particular, notoriously so, for the
case, $n_{1}=30$, $n_{2}=7$ --- but see sec.~\ref{restricted}).
 This instability may be explainable by the 
fact that  the volume element
of the maximal monotone metric, as previously noted, is not normalizable
(to form a probability distribution) over the convex sets of 
$N \times N$ density matrices (in particular, for $N=4$), being
highly singular near the degenerate states (while the volume elements of
the minimal and KMB metrics, though, still singular, are markedly less so,
and are apparently normalizable, extrapolating from the $2 \times 2$
case).
So, one might rely upon either 
the minimal monotone metric or KMB-metric
to provide estimates of the probabilities
of disentanglement (separability) --- as well as simulations of entangled
systems, as ZHSL envisioned.
We believe that estimates based on the minimal monotone
metric should, at least for fine enough grids
and simplicial decompositions, dominate estimates based on any other
member of the {\it continuum}
 of monotone metrics. Following the arguments of Petz and
Sudar \cite{petzsudar}, we contend that any estimates not based
on such metrics (such as the 
``over-parameterized'' results of ZHSL
 \cite{zycz1} --- cf. \cite{boya}  --- and
those of sec.~\ref{semiclassical} here) fail to meet
 certain natural requirements
and should, thus, be taken {\it cum grano salis}.

Let us also note a specific relation between the minimal monotone 
(Bures) metric
and the results of ZHSL. The {\it scalar curvature} of this metric has recently
been shown to attain its {\it minimum} --- $ { (5 N^{2} -4) 
(N^{2}-1) \over 2}$ --- for the totally mixed $N$-dimensional (tracial) 
 state (corresponding to the
$N \times N$ diagonal density matrix having all its nonzero entries
equal to ${1 \over N}$),
and to diverge on the degenerate states, those not of full rank
 \cite{dittmann2}.
Now, in their analysis, ZHSL concluded
both that all states
in a small enough neighborhood of the totally mixed state are separable,
and that the ``purer a quantum state is, the smaller its probability of
being separable'' \cite{zycz1}.
Braunstein {\it et al} have given ``a constructive proof that all mixed
states of $N$ qubits in a sufficiently small neighborhood of the maximally
mixed state are separable'' \cite{braunetal}, while Vidal and Tarrach have
also reached the same conclusion \cite{vidal}.

\subsection{Reanalysis of the anomalous $n_{1} =30$, $n_{2} =7$ case based on
truncation of states near to degeneracy} \label{truncate}

We have also considered the possibility of introducing a third parameter
of choice, that is $\det{\rho}$ --- in addition to $n_{1}$ and
 $n_{2}$ --- into our computations.
It would control the level of degeneracy   below which we reject
for further consideration (due to the singular behavior of
the volume elements of the monotone metrics), the (nearly degenerate)
 density matrices our explicit enumeration
method of sec.~\ref{quantum} generates. (This third parameter has been
implicitly zero in sec.~\ref{quantum}.)
We have, in fact, conducted three additional analyses for the case
$n_{1} =30$, $n_{2} =7$, for which we previously obtained results
of a peculiar nature
(Table~\ref{table1}). The largest possible value the determinant of a $4 \times 4$
density matrix can possess is $ ({1 \over 4})^{4} = {1 \over 256} 
= .00390625$.
In the first analysis, we rejected all those density matrices with
determinants less than ${1 \over 256} \cdot 10^{-4}$, in the second, ${1
 \over 256} \cdot 10^{-3}$ 
and in the third, ${1 \over 256} \cdot 10^{-2}$.
We report these results in Table~\ref{table2}.

It appears then (based on the smallest 
 nonzero threshold, that is ${1 \over 256} \cdot 10^{-4}$)
 that the previously reported (zero-threshold) anomalous behavior
for the maximal and KMB-monotone metrics
was attributable to some set (the number of which we are not
certain) of near-degenerate states 
which, in fact,  passed the partial transposition test for separability.

\subsection{Analyses based on {\it randomized} searches} \label{restricted}

In the previous quantum-theoretic statistical analyses of this section,
 we employed  systematic explicit enumeration methods
to generate $4 \times 4$ density matrices, which we then tested for
separability.
We embarked on such a course after initial computations indicated that it was
extremely difficult to locate the four-by-four density matrices
(in the ambient fifteen-dimensional parameter space) using {\it random} search
methods, in the fashion of ZHSL \cite{zycz1} and sec.~\ref{semiclassical}
of this paper.
Nevertheless, at a later point, we chose to intensively pursue such
a strategy.

In almost
eight hundred and fifty million 
{\it ab initio} searches, we succeeded in obtaining (only) 
sixty-one density matrices --- of which,
twelve  turned out to be separable.
Realizing  that the  ``hit-rate''  would be enhanced if instead of searching
for possible off-diagonal entries in the circle of radius one-half in the
complex plane, we also conducted analyses (though at the risk of
introducing possible biases) based on
radii of one-third and one-fourth, as well
(and also, in a supplementary analysis, five-twelfth).
 The  results are reported in
Table~\ref{tablerandom}.
(As in sec.~\ref{semiclassical},
 standard deviations were {\it not} determined, so no
specific assessment of the number of significant digits in
the probabilistic results is immediately
available.)
They are, then, arguably, generally consistent with the sets of
smaller probabilities reported in Table~\ref{table1}, in particular,
for those based on the largest number of generated density matrices
(corresponding to the bottom rows of the table),
in which we naturally repose the greatest confidence.

\subsection{Probabilities as a function of the participation ratio}

\begin{figure}
\centerline{\psfig{figure=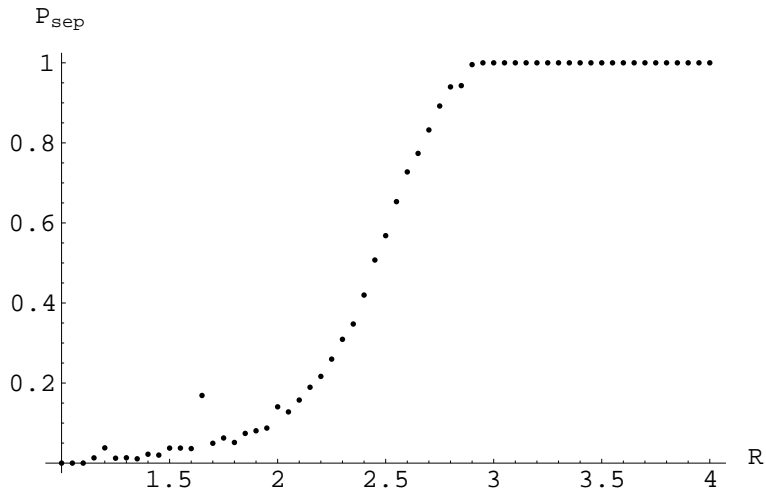}}
\caption{Conditional probability --- based on the minimal monotone
(Bures) metric --- for $N = 2 \times 2$ of finding a separable state, given a 
certain 
range (of width .05) of the
participation ratio $R$, for the scenario $n_{1} = 22, n_{2}=10$}
\label{listplotmin}
\end{figure}

\begin{figure}
\centerline{\psfig{figure=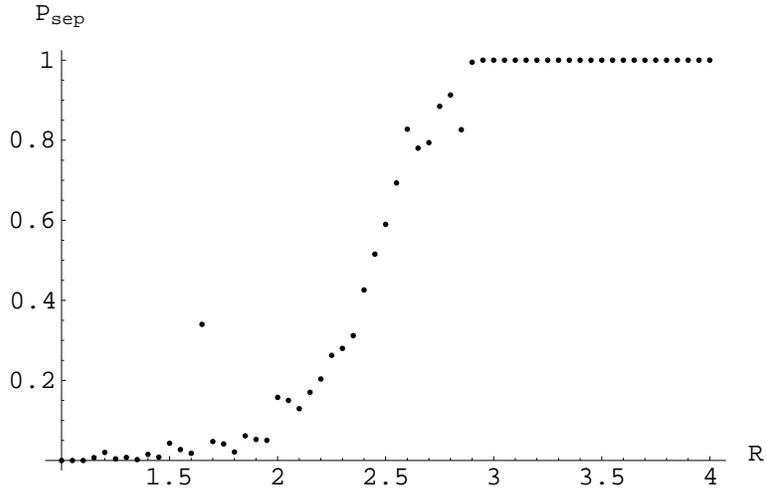}}
\caption{Conditional probability  --- based on the KMB-metric for
$N = 2 \times 2$ --- of
 finding a separable state,
 given a certain range (of width .05) of the
participation ratio $R$, for the scenario $n_{1} =22, n_{2} =10$}
\label{listplotKMB}
\end{figure}

In Fig.~\ref{listplotmin}, we show (using bins of width .05), relying upon
the analysis for the $2 \times 2$ case $n_{1} = 22$, $n_{2} = 10$,
 the 
conditional probability ($P_{sep}$)
 of separability based on the minimal monotone metric, 
 for a given participation ratio $R$ (defined as
 the reciprocal of the trace of the square of the
density matrix \cite[eq. (17)]{zycz1}). In
 Fig.~\ref{listplotKMB},
we show its  counterpart based on  the KMB-metric. (These two figures --- both
having an unexplained ``anomalous blip'' in the interval 
 [1.65, 1.7] --- are the
monotone metric 
analogues of Fig. 2(b) of \cite{zycz1}. It is encouraging, however,
that the ``blip'' does not seem to appear in analogous plots for
other values of $n_{1}$ and $n_{2}$. The value of $R$,
in the $N = 2 \times 2$ case,  must lie between
1 and 4. If $R \geq 3$, the density matrix must be separable
\cite[eq. (18)]{zycz1}.)
\.Zyczkowski \cite{zyczlatest}, drawing upon a long list of open problems
he presents, considers one of the ``most relevant''
to be the question of ``whether the dependence of 
the conditional probability on the participation ratio, obtained for product
measures, holds also for the measures based on the monotone metrics''.
He has hypothesized the existence of
 certain universal/metric-independent features in this regard, that is,
he proposes
that all ``reasonable'' metrics should yield similar such plots.
(He has, in fact, superimposed the figures here upon those previously
generated by him, and found a strong degree of resemblance between them.)

\begin{figure}
\centerline{\psfig{figure=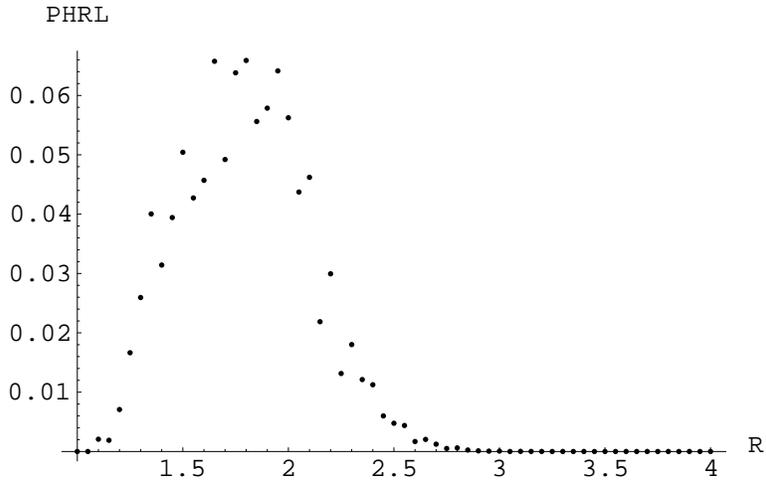}}
\caption{Probability --- based on the minimal monotone (Bures)
 metric --- for $N = 2 \times 2$ of finding a 
 quantum state (whether separable or not),
 given a certain range (of width .05) of
the  participation ratio
$R$, for the scenario $n_{1} = 22, n_{2 }= 10$}
\label{otherplotmin}
\end{figure}

\begin{figure}
\centerline{\psfig{figure=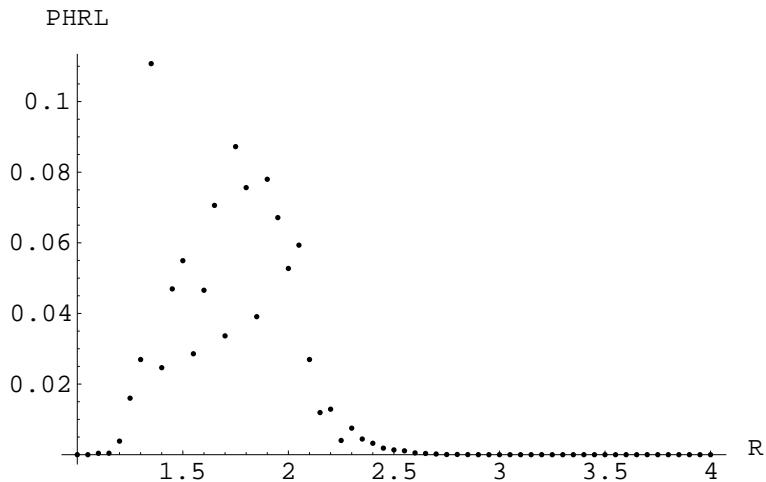}}
\caption{Probability --- based on the KMB-metric -- for $N = 2 \times 2$ of
 finding a 
 quantum state (whether separable or not), given a certain
range (of width .05) of the participation ratio
$R$, for the scenario $n_{1} = 22, n_{2} = 10$}
\label{otherplotKMB}
\end{figure}

In Figs.~\ref{otherplotmin} and \ref{otherplotKMB}, we show 
(again for the case $n_{1} = 22, n_{2} = 10$ of Table~\ref{table1})
 the probabilities
of our generating a density matrix (either separable or inseparable) based
on the minimal monotone and KMB-metrics, respectively.
(These are the monotone metric counterparts of Fig. 2(a) of \cite{zycz1}.
Since we find more probability concentrated at smaller values of $R$
than did ZHSL,
these two figures help us to understand why we obtain smaller 
{\it overall} probabilities of separability ---  in particular,
 less than .5 in the
$N = 2 \times 2$ case --- than their ``surprising''
 result of $0.632 \pm 0.002$.)

\subsection{The case of $2 \times 3$ systems} \label{quanttwobythree}

In Table~\ref{early}, we report initial  findings (of an, unfortunately,
rather unstable nature) for the
$2 \times 3$ systems --- parallel to those given in 
Table~\ref{table1} for the $2 \times 2$ systems.
(We note that the probability of separability obtained by
ZHSL \cite{zycz1} for the $2 \times 3$
systems was $.384 \pm .002$ and, with the 
alternative use of the Jeffreys'
prior in sec.~\ref{semi2by3}, .122.)
There are now six diagonal entries 
(associated with $n_{1}$) and fifteen pairs of off-diagonal entries
(associated with $n_{2}$)
to consider,
so computational demands are substantially increased.
We, of course, expect the probabilities  of separability to be 
{\it less} than the
corresponding ones in the $2 \times 2$ case
reported in Table~\ref{table1}, and this 
is certainly the case for the most extensive analysis
($n_{1} = 20$, $n_{2} = 8$).

The 
somewhat
counterintuitive observation that for $n_{1} = 14$, the choice of $n_{2} = 8$
leads to many more generated density matrices than for $n_{2} = 9$, is
comprehensible in that only 
for even values of $n_{2}$ are no nonzero lower bounds
placed on the possible absolute values of off-diagonal entries.
We were not able for any $n_{2} = 7$ scenario --- due to
memory limitations --- to find a large enough
$n_{1}$, for which any density matrices at all were generated.
We also possess no immediate explanation for the equality of the
three
$p$'s and $d$'s for the three  cases involving $n_{2} = 9$.
(The relatively large probabilities for the scenario, $n_{1} = 24, n_{2} = 6$
may be attributable to
 a ``number-theoretic'' effect, given that 24 is exactly divisible
by 6.)

A randomization approach, such as we pursued in sec.~\ref{restricted}
for the $2 \times 2$ systems, would clearly yield even fewer
density matrices for a given number of
independent searches than there (Table~\ref{tablerandom}).

\section{CONCLUDING REMARKS} \label{Concluding}

We have presented in this study, two forms of evidence 
(one essentially classical and the other, quantum-theoretic in nature)
that the specific choice of (product) 
measure of ZHSL \cite{zycz1} led them to substantially
overestimate the extent to which quantum systems --- in particular,
for $N = 2 \times 2$ and $2 \times 3$ --- should be
considered to possess, in some natural {\it a priori} sense, the
property of separability or disentanglement.
The preponderance of evidence adduced indicates that the probability
of separability for the $N = 2\times 2$
systems, based on the minimal monotone (Bures)
 metric is  no greater than eleven percent --- and if one views 
 the evidence somewhat less
conservatively,  perhaps less than ten percent.
In turn,
estimates founded on any other member of the continuum
of monotone metrics should be lower still. For instance, for
the Kubo-Mori/Bogoliubov metric \cite{petztoth,petzkubo1,balian},
 an estimate of nine percent would
seem conservatively high. Apparently, the 
 {\it maximal} monotone metric --- the volume element of
which possesses a high degree of singularity 
on the degenerate states ($\det{\rho} = 0$), associated with its
conjectured nonnormalizability over the
fifteen-dimensional convex set of states --- must, in some (perhaps limiting) sense,
furnish a {\it lower} bound on the probability of separability.
This bound would have to be strictly greater than zero, if the 
related arguments
made in 
\cite{zycz1,braunetal,vidal,zyczlatest}, in fact, apply.

Let us also note that in continuing work, pertaining to \cite{kratt},
we have found a 
quite interesting distinguished role (that of
 yielding both the {\it minimax} and {\it maximin}
in universal quantum coding) for a monotone metric that has not
apparently previously been noted. Its associated Morozova-Chentsov
function,
\begin{equation}
e ({{\lambda_{i}}^{\lambda_{i}}}/{\lambda_{j}}^{\lambda_{j}})^{1 \over
{\lambda_{j} -\lambda_{i}}},
\end{equation}
 is simply the reciprocal of the
{\it exponential} or {\it identric}
 mean \cite{qi} of $\lambda_{i}$ and $\lambda_{j}$.
The behavior of the related monotone metric appears to be quite close
to that of the minimal (Bures) monotone metric.

The need for additional  computational
 work (possibly utilizing supercomputers) is indicated,
 in regard to what we contend are the theoretically superior
(properly parameterized)
quantum-theoretic analyses of sec.~\ref{quantum},  in order
 to more closely pinpoint estimates. Such analyses could be
 based on finer simplicial decompositions 
for the trial diagonal entries (that is, higher values of
$n_{1}$) and finer  square grids
for the trial off-diagonal entries (that is, higher values of
$n_{2}$),  than those
reported in sec.~\ref{quantum}, and/or possibly randomization procedures,
as in sec.~\ref{restricted}.

At several points in this paper, we have indicated that the analyses of
ZHSL \cite{zycz1} were ``over-parameterized'', in that $N^{2} + N -1$
parameters were  employed, rather than $N^{2} -1$, as is clearly most
natural for the $N \times N$ density matrices (which form an
$N^{2} -1$-dimensional convex set).
However, in a personal communication, M. Byrd has asserted that this
bothersome  feature could be avoided, since $N-1$ Euler angles (in addition
to the
``phase'', as is well known) can be seen to, in fact,  vanish in the 
ZHSL-type representation \cite[eq. (25)]{zycz1} of a density matrix
in the product form $U^{'} D U$.
Byrd  has been able to explicitly show this in the case $N=3$, based on the
Euler angle parameterization given in \cite{byrd} (the  angles $c$ and
$\phi$ vanishing) and contends that analogous phenomena must hold
for $N > 3$, as well (cf. \cite{boya}).
(This vanishing does not appear to occur with the
particular
Euler angle 
parameterizations [associated with Hurwitz]
 used by ZHSL, given in \cite{zycz2,zyczpozniak}.
Byrd suggests that this is because the ``diagonal matrices that make up
the maximal torus'' do not appear on the end, while if they did, they
would commute with the diagonalized density matrix.)
This highly interesting line of thought would suggest that the
analyses of ZHSL, 
\.Zyczkowski \cite{zyczlatest} and those of sec.~\ref{semiclassical} here
(but, of course, not those ``properly parameterized'' ones
of sec.~\ref{quantum})
should be repeated in such a more 
parameter-wise economical framework, and the new results
compared with those previously obtained, to see whether any differences
are found (cf. \cite{slatercomment}).

\acknowledgments

I would like to express appreciation to the Institute for Theoretical
Physics for computational support in this research, and to
K. \.Zyczkowski for several comments, as well as providing the simulations
reported at the end of sec.~\ref{semiclassical}, pointing out
to me \cite{antonelli}, and suggesting that it would be of considerable
interest to ``bin'' the results here, thus, leading to the four included
figures.
I am also grateful to M. Byrd for responding to my queries regarding
the issue of ``over-parameterization'' in the work of ZHSL.

\newpage

\begin{table}
\caption{Quantum-theoretic statistical analyses based on the minimal, KMB and
maximal monotone metrics. The parameters $n_{1}$ and $n_{2}$ determine the
resolution of simplicial decompositions and square grids for trial diagonal
and off-diagonal entries, respectively. The variable $p$ denotes the associated
probability of disentanglement and $d$, the averaged degree of entanglement.
The only generated density matrices which have been omitted from consideration
are those which correspond precisely to degenerate
 states, that is, $\det{\rho} = 0$.
The results are tabulated in increasing
 order of the total number of density matrices
generated --- given in the third column.}

\label{table1}
\begin{tabular}{r r | | r r |   l l l | l l l}
$n_{1}$ & $n_{2}$ & $\rho$ & 
 $\rho_{separable}$ &$p_{min}$ & $p_{KMB}$ & $p_{max}$ & $d_{min}$ 
 & $d_{KMB}$ & $d_{max}$ \\
\hline
23 & 7 & 1,340,928 & 356,096 & .111102 & .0873186 & .0846153 &
.18206 & .208022 & .248457 \\
35 & 6 & 1,425,216 & 467,424 & .193939 & .178191 & .0980763 & .250552 & 
.229161 & .154696 \\
30 & 7 & 2,919,680 & 806,400 & .119669 & .17601 & .749588 & .184696
& .170555 & .00940062 \\
45 & 6 & 3,033,084 & 987,484 & .220936 & .224241 & .293623 &
.248278 & .243704 & .144297 \\
4 & 12 & 4,228,817 & 1,634,577 & .249824 & .232509 & .309404 &
.117852 & .118964 & .0886535 \\
19 & 8 & 4,443,408 & 1,284,816 & .147968 & .123283 & .0554999 & .18686
& .184149 & .17713 \\
8 & 10 & 4,645,163 & 1,230,411 & .11469 & .0982187 & .179838 & .18989 &
.192975 & .143422 \\
35 & 7 & 4,673,024 & 1,286,656 & .0927196 & .0594478 & .152368 &
.220529 & .26694 & .156327 \\
13 & 9 & 5,540,864 & 1,341,440 & .0756821 & .0361165 & .00173733 &
.208862 & .252169 & .349782 \\
6 & 11 & 6,161,152 & 1,703,808 & .114669 & .0859676 & .155728
& .187623 & .205257 & .119427 \\
10 & 10 & 7,103,372 & 2,232,836 & .195802 & .188155 & .243187 &
.131533 & .130076 & .103806 \\
23 & 8 & 8,026,372 & 2,286,148 & .109203 & .0747959 & .000112443 &
.24179 & .260264 & .296032 \\
50 & 7 & 13,522,176 & 3,705,472 & .10125 & .072653 & .687396 &
.206883 & .230803 & .0422876 \\
19 & 9 & 16,603,136 & 4,096,000 & .0920722 & .083066 & .0820477
& .186361 & .188775 & .190464 \\
7 & 12 & 27,658,276 & 6,897,940 & .0779479 & .0553514 & .31028 &
.221055 & .256207 & .0959608 \\
35 & 8 & 28,582,224 & 8,104,555 & .140105 & .122355 & .199223
& .22233 & .228248 & .118103 \\
15 & 10 & 29,328,236 & 8,006,017 & .120671 & .0963975 & .198144 &
.194026 & .210798 & .134718 \\
11 & 11 & 36,913,664 & 9,049,600 & .104142 & .0891733 & .164511 &
.177627 & .179959 & .122263 \\
25 & 9 & 37,671,424 & 9,440,000 & .0837076 & .0343513 & .0027555 &
.202037 & .229187 & .198646 \\
8 & 12 & 40,487,643 & 10,493,443 & .109191 & .0873665 & .16817 &
.185359 & .201147 & .134193 \\
27 & 9 & 47,381,504 & 11,798,016 & .0900729 & .0619535 & .169748 &
.197141 & .219559 & .124459 \\
6 & 13 & 47,815,680 & 12,558,464 & .100372 & .0710627 & .112995 &
.19102 & .205754 & .149789 \\
18 & 10 & 52,099,496 & 13,733,736 & .104203 & .0821789 & .0834828 & .205028
& .215253 & .198921 \\
29 & 9 & 58,888,704 & 14,645,792 & .0886606 & .054769 & .00175691
& .190217 & .201052 & .150746 \\
9 & 12 & 58,900,696 & 14,677,208 & .0933132 & .0726818 & .176224 &
.20039 & .218327 & .131068 \\
13 & 11 & 60,453,376 & 14,847,232 & .0848635 & .0569472 & .062366 &
.193203 & .213676 & .208112 \\
19 & 10 & 61,584,896 & 16,090,832 & .0865202 & .052442 & .155109
& .226729 & .245587 & .127568 \\
30 & 9 & 65,276,416 & 16,252,736 & .0892746 & .0685755 & .0839835 &
.20221 & .239333 & .0959485 \\
32 & 9 & 79,412,992 & 19,729,792 & .0856388 & .0562649 & .123084 &
.200964 & .232005 & .141376 \\
21 & 10 & 83,685,188 & 21,982,132 & .0993573 & .062451 & .138054 &
.20348 & .228876 & .114666 \\
15 & 11 & 92,920,832 & 22,811,392 & .0848161 & .0550803 & .147779 &
.193381 & .211392 & .152707 \\
34 & 9 & 94,713,344 & 23,606,336 & .0913722 & .0638694 & .131152 &
.197612 & .230137 & .2171 \\
22 & 10 & 96,084,402 & 25,272,244 & .106789 & .0808784 & .0769463 &
.205559 & .23114 & .187346 \\
\hline
\end{tabular}
\end{table}

\begin{table}
\caption{Reanalyses of anomalous results 
(Table~\ref{table1}) for $n_{1} = 30$, $n_{2} =7$, using varying
thresholds of degeneracy, as indexed by 
 $\det{\rho}$, below which the generated 
density matrices are rejected from further consideration.}
\label{table2}
\begin{tabular}{r || r r | l l l | l l l}
  threshold  on $\det{\rho}$
 &  $ \rho $ & $ \rho_{separable} $ & $ p_{min} $ & $ p_{KMB} $
 & $ p_{max} $
& $ d_{min} $ & $ d_{KMB} $  & $ d_{max} $ \\
\hline
0 & 2,919,680 & 806,400 & .119669 & .17601 & .749588 & .184696 
& .170555 & .00940062 \\
${1 \over 256} \cdot 10^{-4}$ & 2,913,536
 & 801,792 & .102275 & .0664226 & .00698098
& .189535 & .201332 & .0983051 \\
${1 \over 256} \cdot 10^{-3}$ & 2,856,704 &
 796,672 & .123362 & .108123 & .386157 &
.180807 & .191661 & .101308 \\
${1 \over 256} \cdot 10^{-2}$
 & 2,381,312 & 724,864 & .180938 & .171253 & .260637 &
.14053 & .140938 & .101575 \\
\hline
\end{tabular}
\end{table}

\begin{table}
\caption{Results of {\it random} searches of the fifteen-dimensional parameter
space, using differing  values of the radius ($r$) of the circle in the complex
plane
centered at (0,0),
 from which the possible off-diagonal entries of the $4 \times 4$
density matrices are chosen.}
\label{tablerandom}
\begin{tabular} {r |r  | |  r r | l l l | l l l}
$r$ & searches &
$\rho$  & $\rho_{separable}$ & $p_{min}$ & $ p_{KMB} $ & $ p_{max}$ & $
 d_{min} $  & $ d_{KMB} $  & $ d_{max} $ \\
\hline
1/2 & 847,500,000 & 49 & 12 & .0711773 & .0548709 & .0269786
& .197979 & .2155 & .252218 \\
5/12 & 387,900,000 & 175 & 47 & .098809 & .0408168 & .00218239
& .1736 & .18024 & .194678 \\
1/3 & 351,700,000 & 2,438 & 619 & .0853483 & .0654313 & .180815 & .201774 
& .22259 & .147024 \\
1/4 & 74,300,000 & 15,701 & 3,912 & .0846071 & .0364071 & .00148944 &
.17873 & .20229 & .290831 \\
\hline
\end{tabular}
\end{table}

\begin{table}
\caption{Analogues for the $2 \times 3$ systems of the results of
Table~\ref{table1}}
\label{early}
\begin{tabular}{r r | | r r | l  l l | l l l}
$n_{1}$ & $n_{2}$ & $ \rho $ & $ \rho_{separable} $ & $p_{min} $ & $p_{KMB} $
& $p_{max}$  & $ d_{min} $ & $ d_{KMB} $ & $ d_{max} $ \\
\hline
8 & 8 & 7,581 & 5,205 & .643091 & .654085 & .678443
& .0254769 & .024673 & .0230053 \\
11 & 6 & 35,268 & 9,828 & .000171812 & .000192414 & .000212583 &
.592094 & .592009 & .591929 \\
14 & 6 & 149,607 & 59,727 & .368812 & .425491 & .46159 
& .147718 & .118212 & .10094 \\
15 & 6 & 158,522 & 59,185 & .147933 & .152787 & .128344
& .299006 & .281777 & .235139 \\
13 & 9 & 245,760 & 8,448 & .034375 & .034375 & .034375 &
.185994 & .185994 & .185994 \\
9 & 8 & 235,616 & 65,840 & .234146 & .262858 & .266667 &
.0803952 & .0654659 & .0637942 \\
15 & 9 & 245,760 & 768 & .003125 & .003125 & .003125 & .182635 &
.182635 & .182635 \\
14 & 9 & 368,640 & 9,600 & .0260417 & .0260417 & .0260417 &
.188924 & .188924 & .188924 \\
16 & 6 & 370,479 & 131,055 & .00584908 & .0172682 & .359634
& .581471 & .558051 & .119815 \\
17 & 6 & 557,304 & 164,264 & .000248636 & .000243731 & .000240557
& .59097 & .591328 & .591453 \\
10 & 8 & 579,186 & 150,090 & .173755 & .174011 & .174992 &
.184237 & .194057 & .195405 \\
12 & 8 & 1,352,182 & 303,022 & .0463408 & .0872936 & .19522 
& .383656 & .307713 & .117593 \\
11 & 8 & 1,593,588 & 520,068 & .299328 & .309032 & .326253
& .0995446 & .0968566 & .0940759 \\
22 & 6 & 2,875,965 & 974,818 & .0160034 & .125932 &  .0721558 & 
.562473 & .374128 & .259626 \\
24 & 6 & 3,870,989 & 1,360,213 & .192395 & .19933 & .314609 &
.292782 & .268782 & .140023 \\
13 & 8 & 4,408,872 & 881,397 & .0551892 & .0659333 & .256791 &
.294217 & .259817 & .0915806 \\
14 & 8 & 6,073,071 & 882,781 & .000948857 & .000644427 & .00952389
& .61686 & .476343 & .359429 \\
16 & 8 & 7,373,379 & 1,609,392 & .103614 & .121313 & .203796 &
.204274 & .183523 & .119163 \\
26 & 6 & 7,696,926 & 2,514,130 & .1132 & .111881 & .0999924 & .368272
& .289389 & .229179 \\
15 & 8 & 15,603,746 & 2,109,650 & .000899908 & .00162305 & .00114651 &
.63942 & .599576 & .461553 \\
17 & 8 & 19,413,528 & 3,311,159 & .054504 & .0562656 & .0398168
& .323892 & .298912 & .259056 \\
18 & 8 & 29,075,408 & 5,061,131 & .081246 & .0737912 & .0958513 &
.2304 & .215317 & .181838 \\
20 & 8 & 45,002,652 & 7,755,473 & .00401427 & .0131804 & .0525791 &
.606695 & .487055 & .206391 \\
\hline
\end{tabular}
\end{table}

\listoftables

\listoffigures

\end{document}